\documentclass[conference]{IEEEtran}
\IEEEoverridecommandlockouts
\usepackage{hyperref}
\usepackage{url}

\usepackage{placeins}

\usepackage{cite}
\usepackage{amsmath,amssymb,amsfonts}
\usepackage{algorithmic}
\usepackage{graphicx}
\usepackage{textcomp}
\usepackage{xcolor}
\def\BibTeX{{\rm B\kern-.05em{\sc i\kern-.025em b}\kern-.08em
    T\kern-.1667em\lower.7ex\hbox{E}\kern-.125emX}}
\begin{document}

\title{SDXL Finetuned with LoRA for Coloring Therapy: Generating Graphic Templates Inspired by United Arab Emirates Culture\\

}

\author{\IEEEauthorblockN{Abdulla Alfalasi, Esrat Khan, Mohamed Alhashmi, Raed Aldweik, Davor Svetinovic}
\IEEEauthorblockA{Department of Computer Science, Khalifa University, UAE}
}

\maketitle

\begin{abstract}
A transformative approach to mental health therapy lies at the crossroads of cultural heritage and advanced technology. This paper introduces an innovative method that fuses machine learning techniques with traditional Emirati motifs, focusing on the United Arab Emirates (UAE). We utilize the Stable Diffusion XL (SDXL) model, enhanced with Low-Rank Adaptation (LoRA), to create culturally significant coloring templates featuring Al-Sadu weaving patterns. This novel approach leverages coloring therapy for its recognized stress-relieving benefits and embeds deep cultural resonance, making it a potent tool for therapeutic intervention and cultural preservation. Specifically targeting Generalized Anxiety Disorder (GAD), our method demonstrates significant potential in reducing associated symptoms. Additionally, the paper delves into the broader implications of color and music therapy, emphasizing the importance of culturally tailored content. The technical aspects of the SDXL model and its LoRA fine-tuning showcase its capability to generate high-quality, culturally specific images. This research stands at the forefront of integrating mental wellness practices with cultural heritage, providing a groundbreaking perspective on the synergy between technology, culture, and healthcare. In future work, we aim to employ biosignals to assess the level of engagement and effectiveness of color therapy. A key focus will be to examine the impact of the Emirati heritage Al Sadu art on Emirati individuals and compare their responses with those of other nationalities. This will provide deeper insights into the cultural specificity of therapeutic interventions and further the understanding of the unique interplay between cultural identity and mental health therapy.
\end{abstract}


\section{Introduction}

In recent years, the fusion of technology and traditional cultural motifs has opened new avenues in therapeutic applications, especially the intersection of advanced machine learning techniques with cultural elements. Leveraging sophisticated machine learning capabilities, the research unveils an effective, computationally fast, and sustainable method for crafting intricate and culturally pertinent coloring templates. Recognized for its stress-relieving benefits, coloring therapy gains a dimension of cultural engagement by incorporating patterns and symbols that resonate with the United Arab Emirates (UAE) traditions. This blend offers therapeutic advantages and acts as a conduit for cultural expression and preservation. The template generator contributes to the importance of mental wellness, embodying a holistic approach to well-being. 

The findings of \cite{Samuel2022-iy} indicate that incorporating coloring therapy alongside traditional medication and physiotherapy diminishes both mental and somatic anxiety in patients with Generalized Anxiety Disorder (GAD). Additionally, this approach significantly enhances the reduction of self-reported anxiety levels compared to conventional treatment alone. Generalized Anxiety Disorder (GAD) is a prevalent chronic condition, with an annual occurrence rate ranging from 1.9\% to 5.1\% and a lifetime prevalence of 4.1\% to 6.6\% in the general population. It frequently co-occurs with depression, with comorbidity rates between 45.7\% and 70.0\% \cite{charlson2019new}. Research indicates that ongoing and recurrent symptoms of anxiety and depression significantly impact the psychological well-being and life quality of individuals with GAD. 

This impact includes an increased risk of suicide and a reduction in disability-adjusted life years, thereby placing a substantial psychological and economic strain on individuals, families, and society at large. Consequently, identifying more cost-effective treatments tailored to the psychological needs of GAD patients holds considerable importance for clinical practice and societal health \cite{krupnik2021depression}. In another study \cite{doi:10.1177/0276237420923290}, psychological and psychophysiological effects of coloring and drawing are investigated. The anxiety levels, heart rate, respiratory sinus arrhythmia, and skin conductance are monitored throughout the session. The results indicated that both activities effectively reduced anxiety and altered physiological responses, which are associated with higher levels of enjoyment. 

The graphic template generator also showcases the UAE's cultural vibrancy, artfully merged with modern technology. As mentioned in \cite{Al-Yateem2023-os}, in the course of detailed interviews with children and their mothers in UAE, three elements emerged from the participants' accounts. The first notable element is the profound interrelation among culture, religion, and healthcare approaches. A significant number of mothers demonstrated a preference for traditional remedies as an initial response to health issues, attributing this choice to the deep-rooted cultural significance of these practices. 

\begin{figure}[ht!]
    \centering
    \includegraphics[width=\columnwidth]{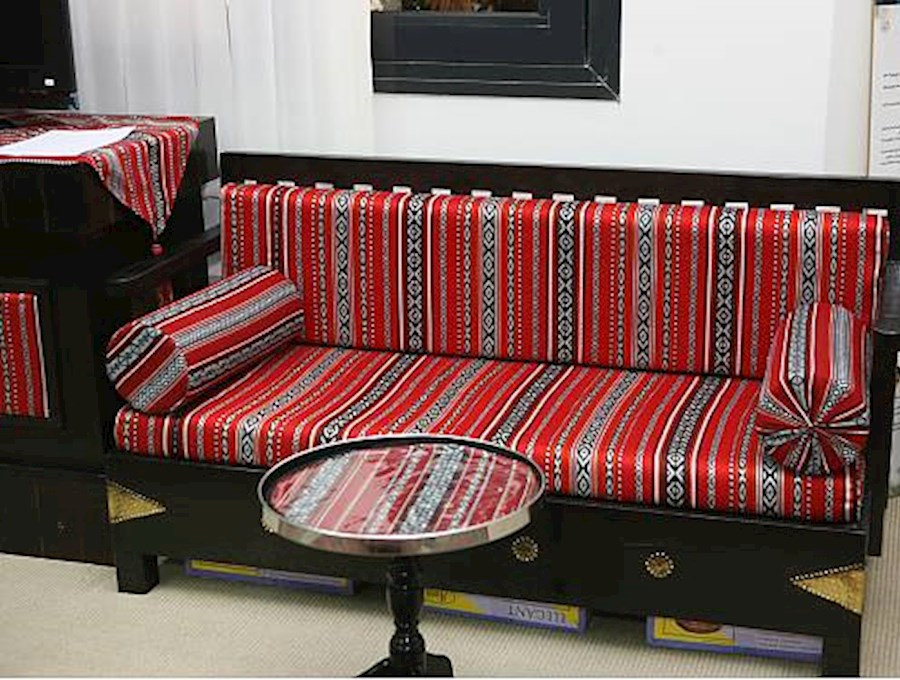}
    \caption{Al-Sadu, a traditional form of weaving practiced by women in the UAE \cite{IntangibleCulturalHeritage2019}.}
    \label{fig:sadu}
\end{figure}

The cultural practices are deemed efficacious and provide spiritual solace, thereby strengthening the participants' cultural identity. It is against this backdrop that the Stable Diffusion XL (SDXL) model is finetuned with Low-Rank Adaptation (LoRA) for the generation of graphic templates inspired by UAE's most valuable cultural weaving patterns called Al-Sadu for coloring therapy in the Emirati society as shown in \autoref{fig:sadu}. Al-Sadu is central to the Bedouin culture in the UAE and is recognized by the United Nations Educational, Scientific and Cultural Organization (UNESCO) \cite{IntangibleCulturalHeritage2019}. This centuries-old Emirati craft was the theme of the $52^{nd}$ Union Day celebrations \cite{Shouk_2023}.

\FloatBarrier

\section{Background and Related Works}

\subsection{The Effectiveness of Chromotherapy on Youth}

Modern healthcare integrates color therapy as a complementary treatment, recognizing its impact on mental and physical well-being. Different colors can alter energy levels, mood, appetite, and decision-making, with some colors energizing and others calming. Color therapy, based on the premise that different hues affect individuals differently, potentially offers benefits like improved blood pressure, mood, stress relief, and even help with disorders like hyperactivity and insomnia. This practice, which varies in interpretation across cultures, works by our brain interpreting light waves as colors, each with unique emotional and mental resonances. 

The application of color therapy in mental health helps professionals understand behavior-influencing mental processes. Different colors elicit specific psychological effects, like red increasing energy and heart rate, orange inspiring creativity, yellow enhancing mood, green promoting relaxation, and blue aiding anxiety relief. Understanding the psychological, emotional, and physical impact of colors, even in educational settings for children, reveals their profound effect across cultures and emotional states. 

This study utilized primary and secondary data to explore the impact of colors on youth's mental states. Youths aged 15-25 are surveyed using a specialized questionnaire assessing their favorite colors, perceptions of chromotherapy, and effectiveness. Additionally, mood-based color charts are used to note their emotional responses. The study aimed to integrate chromotherapy with cognitive behavioral therapy to address negative thoughts and behaviors among youths, focusing on issues like aggression, anxiety, and criminal tendencies. The survey revealed that their moods and views on life influence youths' color preferences before learning about chromotherapy. It highlighted the effectiveness of chromotherapy as a complementary and alternative medicine, with 80\% of youths embracing it for well-being improvement \cite{Kriticka_2023}.

\subsection{Evaluating the Color Preferences for Elderly Depression in the United Arab Emirates}

Older adults are at a heightened risk of depression due to various physical, psychological, and economic transformations. In the United Arab Emirates (UAE), approximately 25.7\% of the elderly population experiences depression. Color therapy has gained recognition as an effective approach to alleviating depressive symptoms in this demographic. This study focused on the impact of color preferences in the living spaces of the elderly at the Seniors’ Happiness Centre in Ajman, UAE, on their quality of life and depression symptoms. 

The research investigated the association between the preferred colors in the residential bedrooms and the symptoms of depression among elderly residents. The methodology involved eliciting physiological and psychological responses from 86 elderly participants to their preferred bedroom colors using color images. These responses were evaluated in a simulated 3D space using a viewing box. The study employed the Geriatric Depression Scale (GDS) and Electroencephalogram (EEG) measurements to explore the connection between color preference and depression. Findings indicated a stronger inclination towards warm colors among the elderly and the need for varied color schemes due to differing visual characteristics in individuals aged 65 and above. No significant disparity was observed in the participants' alpha wave values in the prefrontal cortex, attributed to the individual variability in brain wave signals. A wall color scheme with heightened saturation showed the potential to reduce depressive symptoms. Psychologically, healthy elderly individuals responded positively to monochromatic blue schemes, whereas those with depression responded favorably to contrasting blue-yellow/red schemes. As a result, engaging the five senses through color stimulation, color therapy can promote psychological equilibrium, offering therapeutic benefits and influencing metabolic processes in the body \cite{buildings12020234}.

\subsection{Long Short-Term Memory-Based Music Analysis System for Music Therapy}

Music serves as a medium for expressing individual thoughts and emotions. The discipline of music therapy utilizes a variety of musical activities, including listening, singing, playing instruments, and engaging with rhythm, to stimulate and influence the human brain. Integrating artificial intelligence (AI) into music therapy has led to groundbreaking advancements throughout the entire diagnosis, treatment, and evaluation process. Leveraging AI's capabilities is essential to innovate in music therapy techniques, improve the precision of treatment plans, and broaden the horizons of medical research. This paper introduces a Long Short-Term Memory (LSTM)-based algorithm for generating and classifying multi-voice music data. A system named Multi-Voice Music Generation (MVMG), founded on this algorithm, is presented. MVMG operates in two primary stages. Initially, music data are transformed into MDPI and textual sequence data through an autoencoder model, encompassing the extraction of musical features and the representation of music clips. Subsequently, an LSTM-based music generation and classification model is developed, tailored for use in specific therapeutic contexts. MVMG's efficacy is assessed using our collected datasets, which include single-melody MIDI files and a Chinese classical music dataset. Results indicate that the autoencoder-based feature extractor achieves a peak accuracy of 95.3\%. Furthermore, the LSTM model's average F1-score is 95.68\%, significantly surpassing that of Deep Neural Network (DNN)-based classification models \cite{10.3389/fpsyg.2022.928048}.

\subsection{Affective computing in the context of music therapy: a systematic review}

Music therapy has shown promise in decelerating the progression of dementia, as engaging with music can activate emotional responses, stimulating brain regions linked to memory. The effectiveness of this therapy is heightened when therapists deliver tailored stimuli to each patient, a task that can be challenging. Artificial Intelligence (AI) methodologies offer potential assistance in this area. This paper presents a systematic review of literature on affective computing within the realm of music therapy, with a specific focus on evaluating AI techniques for automatic emotion recognition in Human-Machine Musical Interfaces (HMMI). The review was conducted through an automated search across five major intelligent computing, engineering, and medicine databases, targeting publications from 2016 to 2020. The search parameters were set to include papers with relevant terms in their metadata, titles, or abstracts. From the initial 290 publications identified, 144 were included in the systematic review. This comprehensive analysis allowed for identifying current challenges in automatic emotion recognition and highlighted the potential of this technology in developing non-invasive assistive tools based on HMMI. It also sheds light on the AI techniques currently employed in emotion recognition from multimodal data. Consequently, using machine learning for emotion recognition from various data sources is a crucial strategy to enhance the therapeutic objectives achievable through music therapy \cite{Santana_Lima_Torcate_Fonseca_Santos_2021}.

\subsection{On the use of AI for Generation of Functional Music to Improve Mental Health}

Recent research highlights music's significant impact on physical and mental health, including enhancements in cardiovascular health, decreased dementia prevalence in older populations, and improved overall mental well-being indicators like stress reduction. This paper outlines brief case studies that explore general mental well-being improvements, specifically anxiety and stress reduction, through AI-powered music generation. Active engagement in listening to and creating music, particularly for vulnerable age groups, has proven highly beneficial. Music therapy has demonstrated its efficacy across various applications and age groups. Nonetheless, barriers to music creation include limited access to expertise, materials, and financial constraints. In this study, a machine learning technique for generating functional music, guided by bio-physiological measurements, is outlined. These studies target emotional states at opposite ends of a Cartesian affective space, a dimensional model of emotion ranging from relaxation to fear. Galvanic skin response, indicative of psychological arousal and estimated emotional state, is a control signal in training the machine learning algorithm. This algorithm generates a non-linear time series of musical features for real-time sound synthesis, employing a perceptually informed model of musical feature similarity. The findings indicate a correlation between familiarity and perceived emotional response \cite{10.3389/frai.2020.497864}.

\subsection{Our Contribution}

The emerging field of generative AI, a term often used to describe a new category of technological tools, has sparked significant discussion. Notably, its applications in creating superior artistic content are remarkable, spanning visual arts, concept designs, music, literature, and even extending to video and animation. Diffusion models, for instance, generate images of exceptional quality \cite{9726576}. These innovative technologies are set to fundamentally transform the creative process, offering new and dynamic ways for individuals to imagine and actualize their artistic visions \cite{101126scienceadh4451}.

A notable research gap exists in the application of coloring therapy within a cultural context. Prior studies have predominantly focused on chromotherapy and music therapy, neglecting the impact of incorporating culturally specific content. The effectiveness of coloring therapy may vary significantly when using graphic patterns that resonate with a particular culture compared to those derived from different cultural backgrounds. This aspect of cultural specificity in therapeutic outcomes remains largely unexplored and warrants further investigation. Consequently, this study generates culture-specific patterns for coloring therapy, utilizing the Stable Diffusion XL (SDXL) model, which has been fine-tuned with Low-Rank Adaptation (LoRA). Looking ahead, our future work aims to employ biosignals to assess the level of engagement and effectiveness of color therapy. Specifically, we plan to examine the impact of the Emirati heritage Al Sadu art on Emirati individuals, comparing their responses with those of other nationalities.

The structure of the remainder of this paper is organized as follows. Section \ref{Sec3} provides a detailed exposition of the SDXL base model, integrating the LoRA (Kohya SS) fine-tuner and the corresponding parameter configurations. Following this, Section \ref{Sec4} presents a comprehensive evaluation of the experimental results, showcasing the model's efficacy by generating Al-Sadu patterns. Lastly, Section \ref{Sec5} concludes the key findings and implications of the research with the scope of future work.

\section{Design of Base Model with Low-Rank Adaption of Large Language Model} 
\label{Sec3}

\begin{figure}[b!]
    \centering
    \includegraphics[width=\columnwidth]{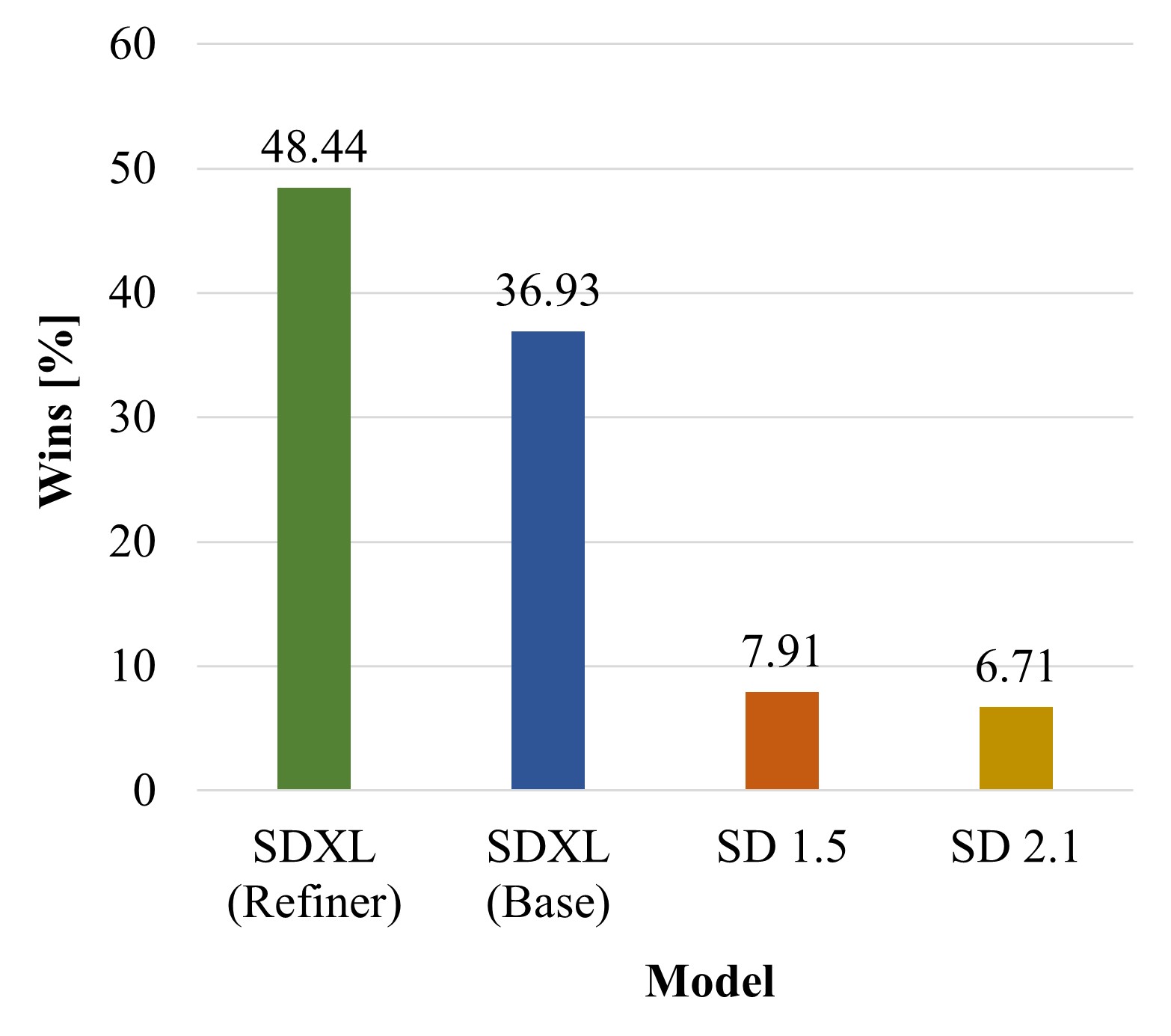}
    \caption{Evaluating user preferences between SDXL and earlier versions of Stable Diffusion, namely 1.5 and 2.1 \cite{podell2023sdxl}.}
    \label{fig:SDchartComparison}
\end{figure}

Recently, there have been significant advancements in deep generative modeling across various domains, including language, audio, and visual media. This report focuses on the visual images aspect, introducing SDXL, an enhanced version of the Stable Diffusion model. SDXL stands out by significantly outperforming previous models, as shown in \autoref{fig:SDchartComparison} and \autoref{tab:tab1}. Its enhanced performance is attributed to a larger UNet-backbone, innovative conditioning techniques without extra supervision, and a diffusion-based refinement process for better visual quality. Additionally, addressing concerns in image creation regarding the opacity of black-box models, which hinders reproducability and innovation, SDXL offers an open model alternative. It provides competitive performance while enabling a more transparent assessment of biases and limitations, crucial for responsible and ethical use \cite{podell2023sdxl}.

\begin{figure*}[ht!]
    \centering
    \includegraphics[width=\textwidth]{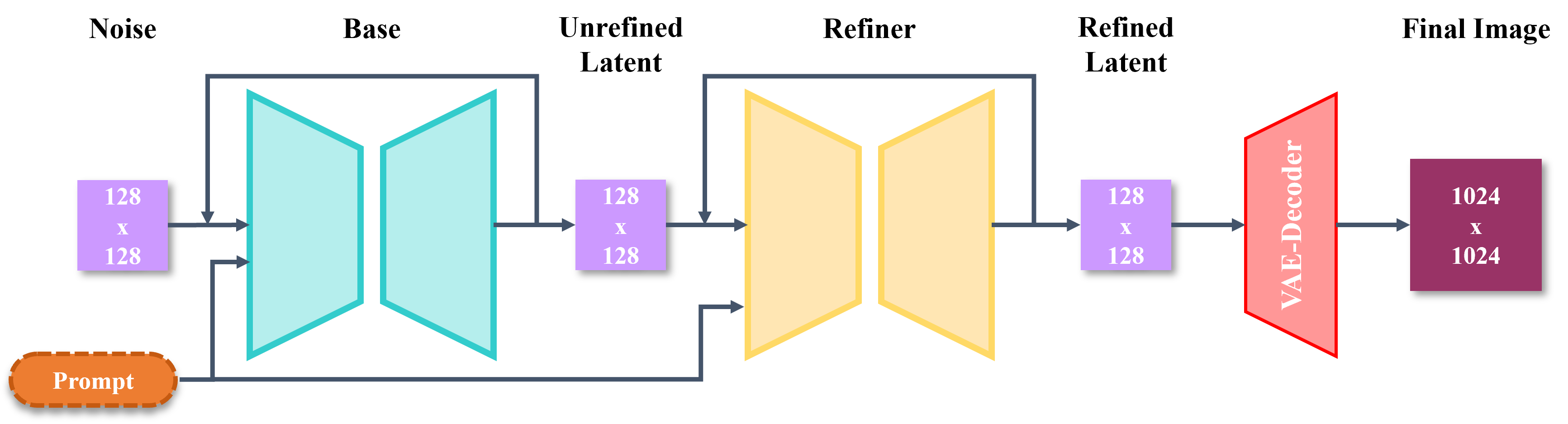}
    \caption{SDXL Model \cite{podell2023sdxl}.}
    \label{fig:SDXLmodel}
\end{figure*}

The convolutional UNet architecture has dominated image synthesis, evolving through enhancements like self-attention, improved upscaling, and cross-attention for text-to-image synthesis. Recent trends include shifting transformer computation to lower-level features in the UNet, as adopted in the SDXL model, which differs from the original Stable Diffusion architecture by distributing transformer blocks heterogeneously within the UNet as shown in \autoref{tab:tab1}. SDXL employs a powerful combination of OpenCLIP ViT-bigG and CLIP ViT-L for text encoding, with the penultimate outputs concatenated along the channel axis. This model, with a 2.6B parameter UNet, also innovates by conditioning on pooled text embeddings from OpenCLIP. Addressing the image size issue in latent diffusion models (LDMs), SDXL conditions the UNet model on the original image resolution, avoiding the shortcomings of discarding low-resolution images or introducing upscaling artifacts. SDXL's unique size-conditioning at the inference stage allows users to set the desired resolution, with the model adapting to resolution-dependent features. This approach also mitigates the issue of blurry samples and poor generalization found in models that discard low-resolution training images. The final SDXL model undergoes a multi-stage training process, incorporating a refined autoencoder and a discrete-time diffusion schedule. It further improves image quality through a refinement stage, employing a noising-denoising process on base model outputs \cite{podell2023sdxl}.

\begin{table}[t!]
\caption{SDXL vs older Stable Diffusion models \cite{podell2023sdxl}.}
\label{tab:tab1}
\resizebox{\columnwidth}{!}{%
\begin{tabular}{|c|c|c|c|}
\hline
\emph{Model}                        & \emph{SDXL} & \emph{SD 1.5} & \emph{SD 2.1} \\ \hline
\textit{\emph{No. of U-Net params}} & 2.6B          & 860M            & 865M            \\ \hline
\textit{\emph{Transfer Blocks}}     & [0, 2, 10]    & [1, 1, 1, 1]    & [1, 1, 1, 1]    \\ \hline
\textit{\emph{\begin{tabular}[c]{@{}c@{}}Channel \\ Multiplication\end{tabular}}} &
  [1, 2, 4] &
  [1, 2, 4, 4] &
  [1, 2, 4, 4] \\ \hline
\textit{\emph{Text Encoder}} &
  \begin{tabular}[c]{@{}c@{}}CLIP ViT-L \& \\ OpenCLIP ViT-bigG\end{tabular} &
  CLIP ViT-L &
  OpenCLIP ViT-H \\ \hline
\textit{\emph{Context Dimension}}   & 2048          & 768             & 1024            \\ \hline
\textit{\emph{\begin{tabular}[c]{@{}c@{}}Pooled Text \\ Embeddings\end{tabular}}} &
  OpenCLIP ViT-bigG &
  Not Available &
  Not Available \\ \hline
\end{tabular}%
}
\end{table}

We utilize a fine-tuning method known as LoRA \cite{hu2021lora}, which allows large base models like SDXL to undergo fine-tuning with a variable rank to achieve the cultural weaving art pattern. In this context, the key advantage of employing a variable rank is the ability to conduct fine-tuning efficiently on a single GPU, making the process more accessible and manageable. The initial step involves gathering images that match the targeted resolution and style. For the Al-Sadu knitting pattern LoRA, around 10 high-quality images were collected, and for the art therapy LoRA, approximately 80 high-quality images of art therapy were gathered. Subsequently, all images are cropped to a 1:1 aspect ratio, a method empirically observed to enhance results in training images for LoRA. To caption the images, the Kohya SS tool \cite{YubinIBM_Anaka_Ma_2023} was used, employing either manual methods or automated techniques like BLIP captioning \cite{li2022blip}; in this case, WD14 captioning was utilized.

Once the dataset is captioned, the images and their corresponding caption text files are used to train the LoRA, employing a base model (SDXL in our instance). The training process took roughly two hours and yielded a LoRA capable of generating images in the desired style. Multiple relevant settings \cite{Ma_2023} are adjusted during the training stage, such as the rank of the LoRA model. A higher rank results in a fine-tuned model that produces images more closely resembling the provided dataset, while a lower rank skews the output images closer to the base model used for fine-tuning. The number of epochs is also altered during training, with an epoch representing a complete forward and backward pass in training. Additionally, the learning rate is modified, which affects the size of weight updates per epoch.

\section{Results of the Finetuning on the Base Model} 
\label{Sec4}

This section visually represents the outcomes achieved through the LoRA fine-tuning methodology. The following examples highlight the diversity and quality of images generated after training with a specific dataset and settings. The generated images exhibit a remarkable adaptation to the datasets' nuances. In the Sadu knitting pattern image, the LoRA model effectively captures intricate details, showcasing its ability to reproduce cultural and artistic elements. Similarly, the art therapy images reflect a rich diversity of styles and shapes, aligning with the varied nature of the art therapy dataset.

\subsection{Al-Sadu Weaving Pattern Images}

\begin{figure}[ht!]
    \centering
    \includegraphics[width=\columnwidth]{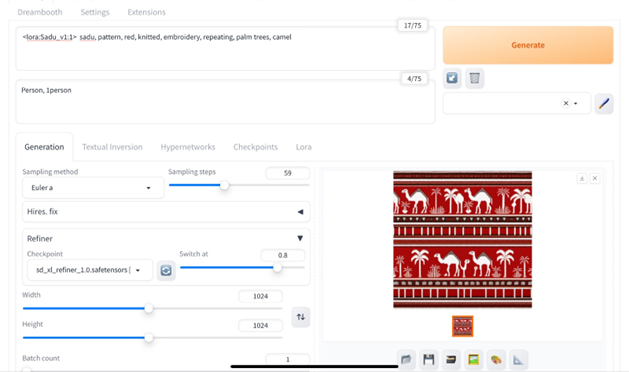}
    \caption{Settings for generating Camel and Palm Tree Al-Sadu.}
    \label{fig:settings1}
\end{figure}

\begin{figure}[ht!]
    \centering
    \includegraphics[width=\columnwidth]{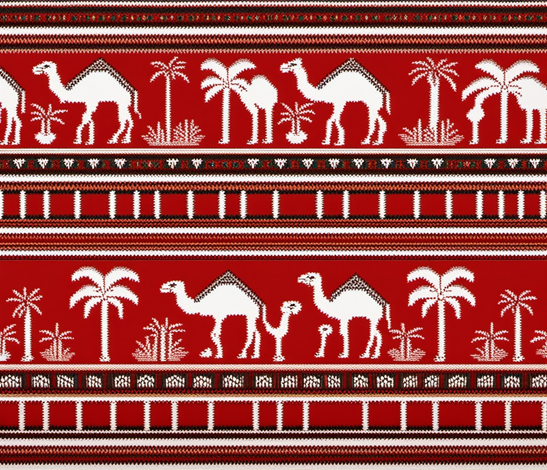}
    \caption{Al-Sadu: Generated Camel and Palm Tree Image 1 with few colors.}
    \label{fig:Sadu1}
\end{figure}

\begin{figure}[ht!]
    \centering
    \includegraphics[width=\columnwidth]{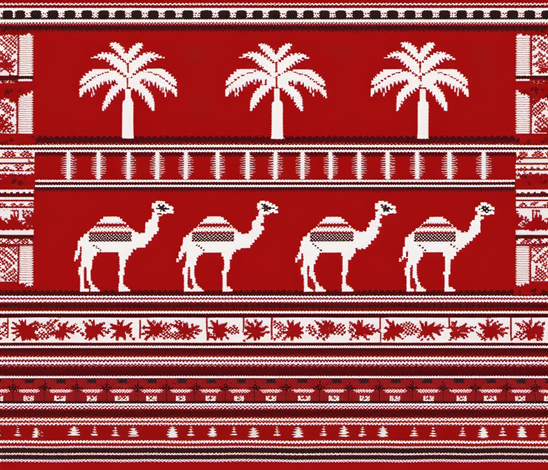}
    \caption{Al-Sadu: Generated Camel and Palm Tree Image 2 with few colors.}
    \label{fig:sadu2}
\end{figure}

\begin{figure}[ht!]
    \centering
    \includegraphics[width=\columnwidth]{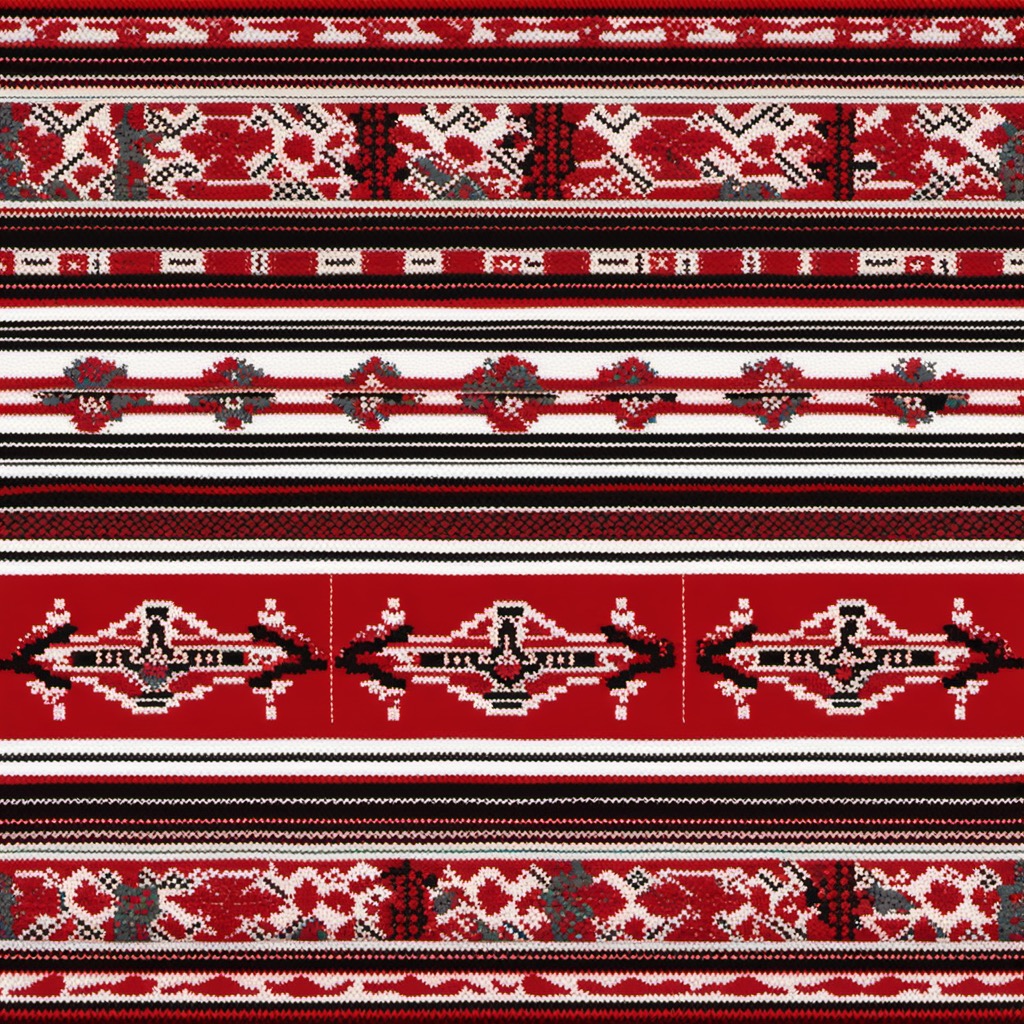}
    \caption{Al-Sadu: Generated Image 3 with few colors.}
    \label{fig:saduImg2}
\end{figure}

\begin{figure}[ht!]
    \centering
    \includegraphics[width=\columnwidth]{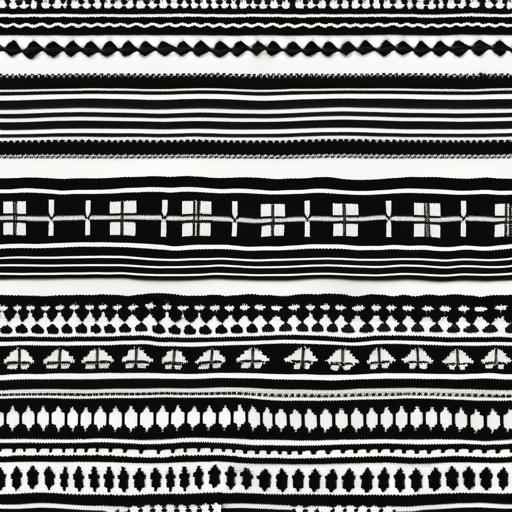}
    \caption{Al-Sadu: Generated Image 4 in black and white.}
    \label{fig:AlSaduColor1}
\end{figure}

\begin{figure}[ht!]
    \centering
    \includegraphics[width=\columnwidth]{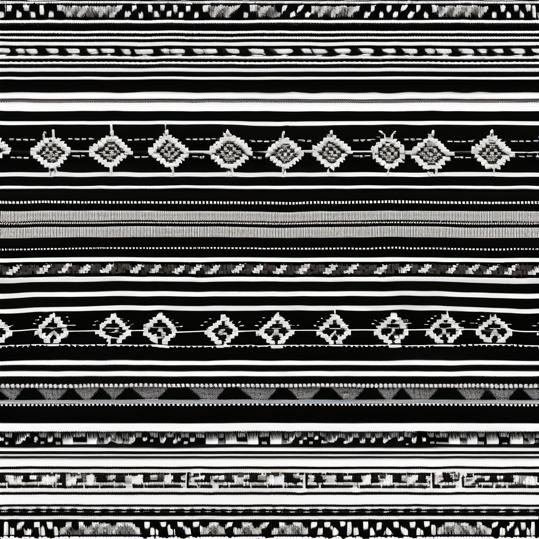}
    \caption{Al-Sadu: Generated Image 5 in black and white.}
    \label{fig:AlSaduColor2}
\end{figure}

\autoref{fig:settings1} displays the GUI with the settings for generating the Camel and Palm Tree Al-Sadu. The sampling method is chosen by Euler 'a' which is the numerical method used for the diffusion process during image generation. The sampling step is set at 59, indicating how many iterations the model runs through to refine the generated image. The specific refiner model integrated is the \textit{sd\_xl\_refiner\_1.0.safetensors}. The dimensions of the generated images are 1024 by 1024. To generate a single image, the batch count is set to 1. \autoref{fig:Sadu1} and \autoref{fig:sadu2} shows the Camel and Palm Tree Al-Sadu generated images with a few colors of the UAE flag in red, green, white and black. Similarly, \autoref{fig:saduImg2} is a pure cultural weaving pattern of Al-Sadu in the UAE flag colors of red, green, white and black. In exploring color therapy through cultural Al-Sadu patterns, \autoref{fig:AlSaduColor1} and \autoref{fig:AlSaduColor2} present generated black-and-white images. These figures demonstrate the intricate complexity of the patterns, with high-resolution imagery ensuring that even the most subtle details are discernible.

\FloatBarrier

\FloatBarrier

\FloatBarrier

\FloatBarrier

\FloatBarrier

\FloatBarrier

\subsection{Other Art Therapy LoRA Images}

\begin{figure}[ht!]
    \centering
    \includegraphics[width=\columnwidth]{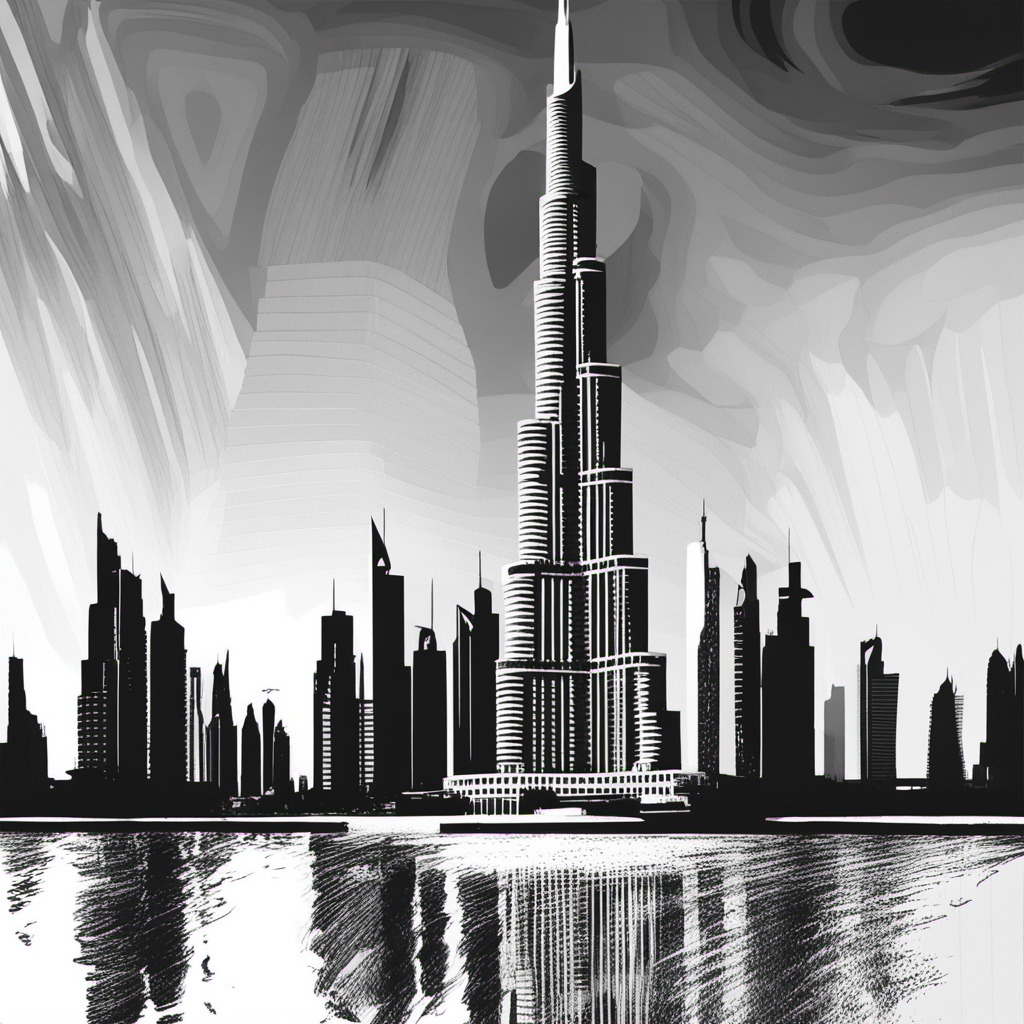}
    \caption{Burj Khalifa image generated.}
    \label{fig:denoiseOrig}
\end{figure}

\begin{figure}[ht!]
    \centering
    \includegraphics[width=\columnwidth]{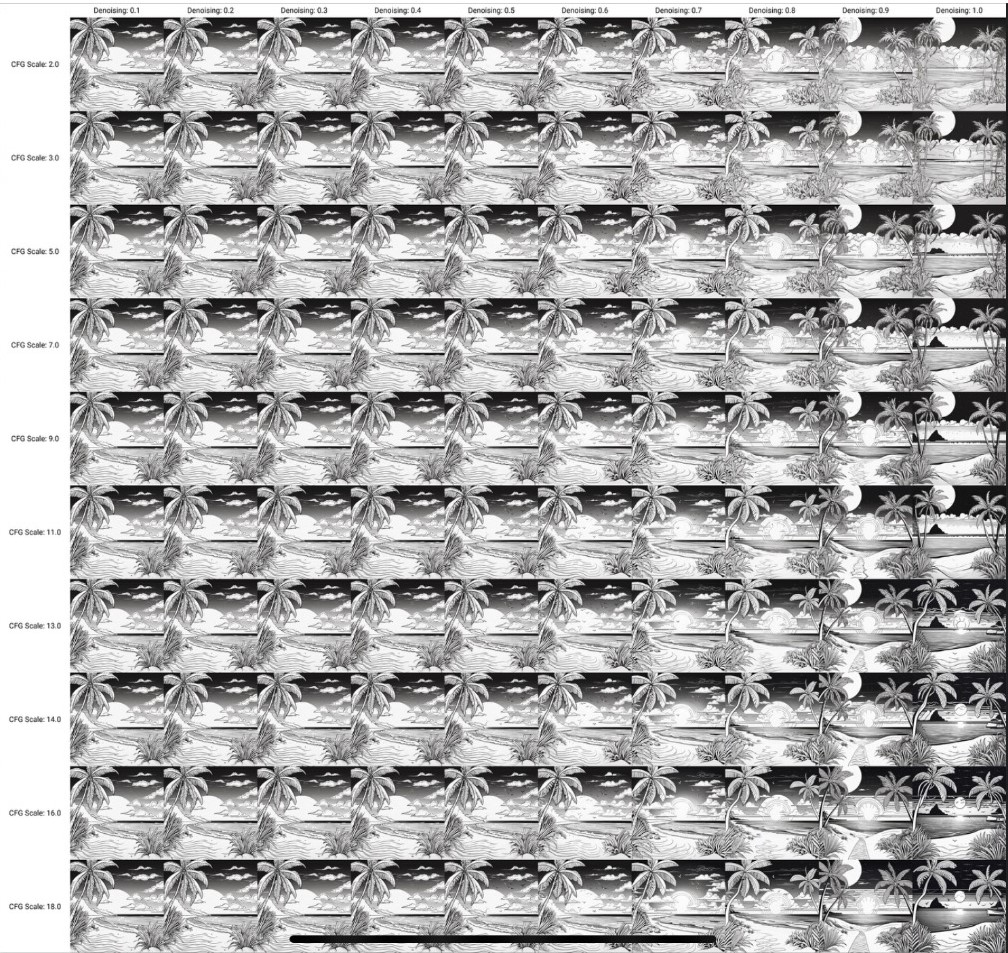}
    \caption{Plot of two parameters changing, CFG scale and denoising factor for the Beach Sunset image.}
    \label{fig:denoising}
\end{figure}

\begin{figure}[ht!]
    \centering
    \includegraphics[width=\columnwidth]{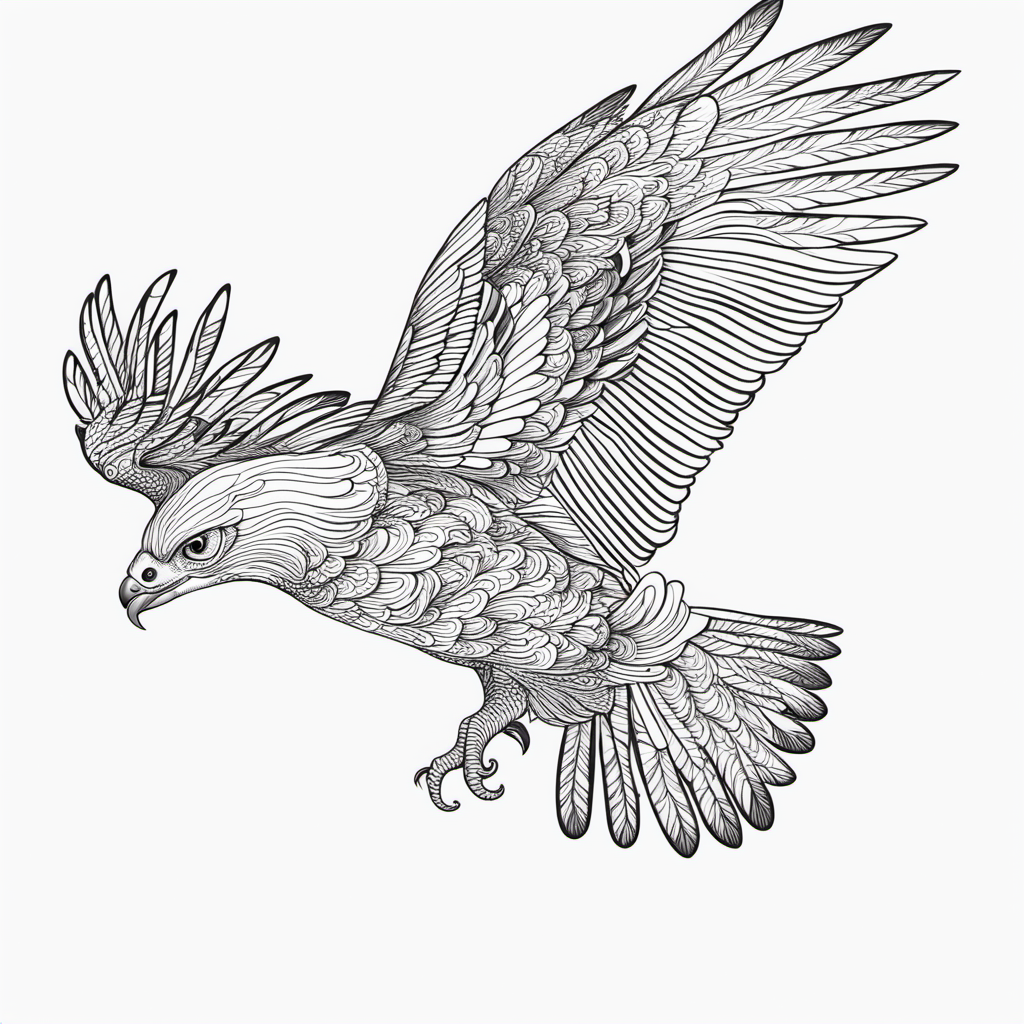}
    \caption{Detailed Falcon generated image.}
    \label{fig:Duck}
\end{figure}

The results illustrate the effectiveness of the LoRA fine-tuning approach in creating a variety of detailed images, reflecting the model's capacity to adjust to different data inputs, as represented by the generated falcon image in \autoref{fig:Duck} and Burj Khalifa in \autoref{fig:denoiseOrig} using a LoRA trained on art therapy images. Using variable rank tuning demonstrates LoRA's practicality for tasks in image generation. Moreover, \autoref{fig:denoising} is the XY plot that provides a clear visual representation of how changes in the CFG scale and denoising factor influence the image synthesis process, offering valuable insights into the fine-tuning mechanics.

\FloatBarrier

\FloatBarrier

\FloatBarrier

\subsection{The Merits of LoRA Fine-Tuning in SDXL Model Optimization}

The LoRA (Kohya SS) fine-tuning method, when integrated with the SDXL model, offers several benefits:

\emph{Resource Efficiency:} LoRA allows for the fine-tuning of large models like SDXL without the need for extensive computational resources. It is designed to be effective even when operating on a single GPU, making it more accessible for researchers and practitioners without high-end hardware.

\emph{Preservation of Pre-trained Knowledge:} LoRA fine-tunes the model by introducing only a small number of additional parameters. This approach retains the pre-trained weights and knowledge of the SDXL model, ensuring that the nuanced understanding captured during its initial extensive training is not lost.

\emph{Fine-grained Customization:} By adjusting the rank in LoRA, fine-grained control over the fine-tuning process is possible. This means the model is more precisely tailored to specific domains or datasets, such as generating culturally significant patterns like Al-Sadu.

\emph{Scalability:} The variable rank characteristic of LoRA provides scalability, allowing the fine-tuning to adapt to the size of the dataset. This scalability makes it suitable for both small and large datasets, providing flexibility in application.

\emph{Rapid Iteration and Development:} Since LoRA requires fewer resources and is efficient in fine-tuning, it supports rapid prototyping and iteration, which is crucial in research and development environments where time and resources are at a premium.

\emph{Quality Retention in Generative Tasks:} In generative tasks, particularly those involving complex patterns and textures such as those found in cultural artifacts, LoRA helps maintain the high quality of the generated images. It ensures that the fine details and subtleties of the original patterns are captured and reproduced effectively.

\emph{Adaptability to New Tasks:} LoRA enables the SDXL model to be quickly adapted to new tasks and datasets. This flexibility is particularly valuable in fields that require the model to generate various outputs, such as diverse art styles or novel scientific visualizations.
\subsection{LoRA Links}
We made the models discussed in this paper publicly available on the Hugging Face platform, the Sadu pattern analysis model is available at \href{https://huggingface.co/aghanim1/sadu}{https://huggingface.co/aghanim1/sadu} and the art therapy model is available at \href{https://huggingface.co/aghanim1/arttherapy}{https://huggingface.co/aghanim1/arttherapy}. These links provide direct access to the models, as well as a provision to generate images, although images generated using Automatic1111 and the model will provide better results.

\section{Conclusion}
\label{Sec5}

With a special focus on the rich and diverse cultural heritage of the United Arab Emirates (UAE), coloring therapy, recognized for its stress-relieving properties, gains an additional layer of significance by integrating culturally resonant patterns and symbols. This integration provides a therapeutic benefit and serves as a medium for cultural expression and preservation. By harnessing the power of advanced machine learning, this research demonstrates a novel method of creating intricate and culturally relevant coloring templates, offering a unique blend of mental wellness and cultural engagement. The resultant templates are therapeutic tools and serve as a testament to the UAE's vibrant culture, showcasing its aesthetic elements through a modern technological lens.

\appendix[Other Images Generated by Our Model]

\begin{figure}[ht!]
    \centering
    \includegraphics[width=\columnwidth]{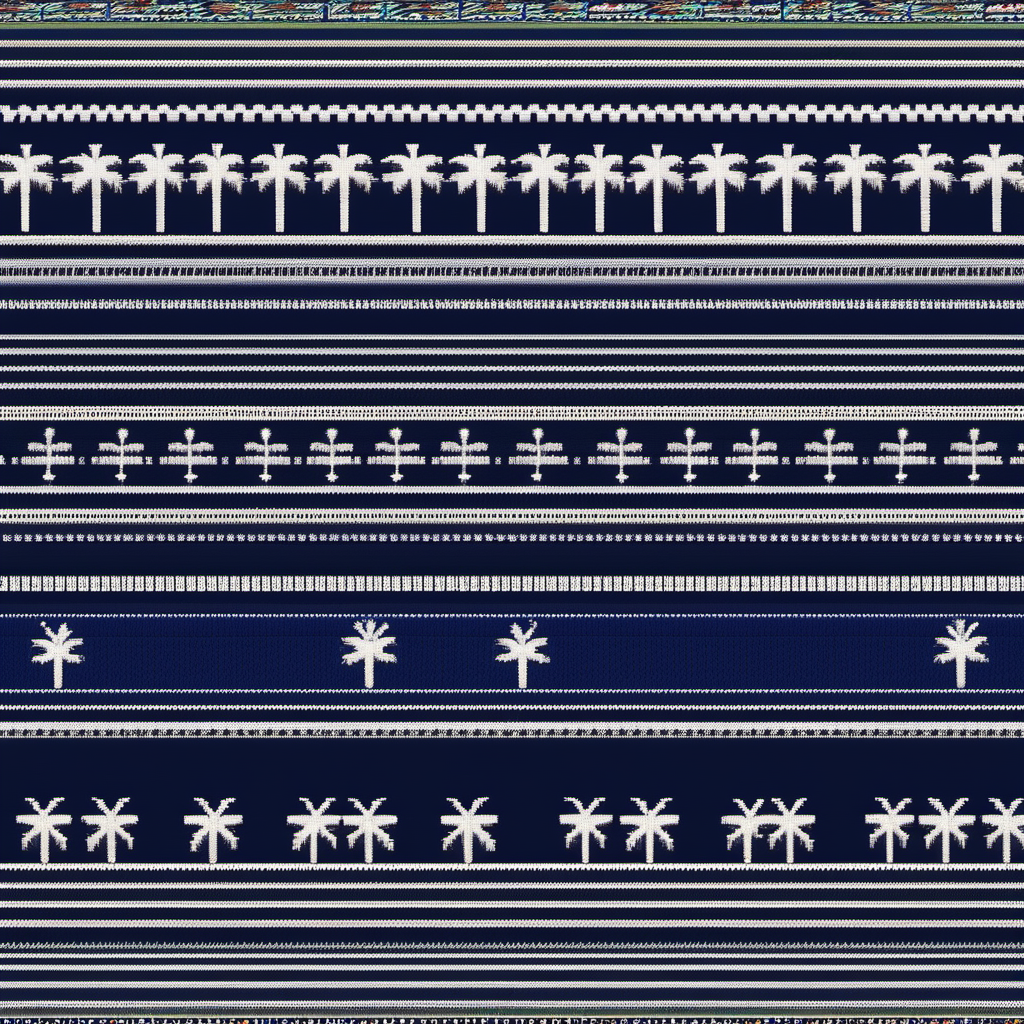}
    \caption{Al-Sadu generated image.}
    \label{fig:00034}
\end{figure}
\FloatBarrier

\begin{figure}[ht!]
    \centering
    \includegraphics[width=\columnwidth]{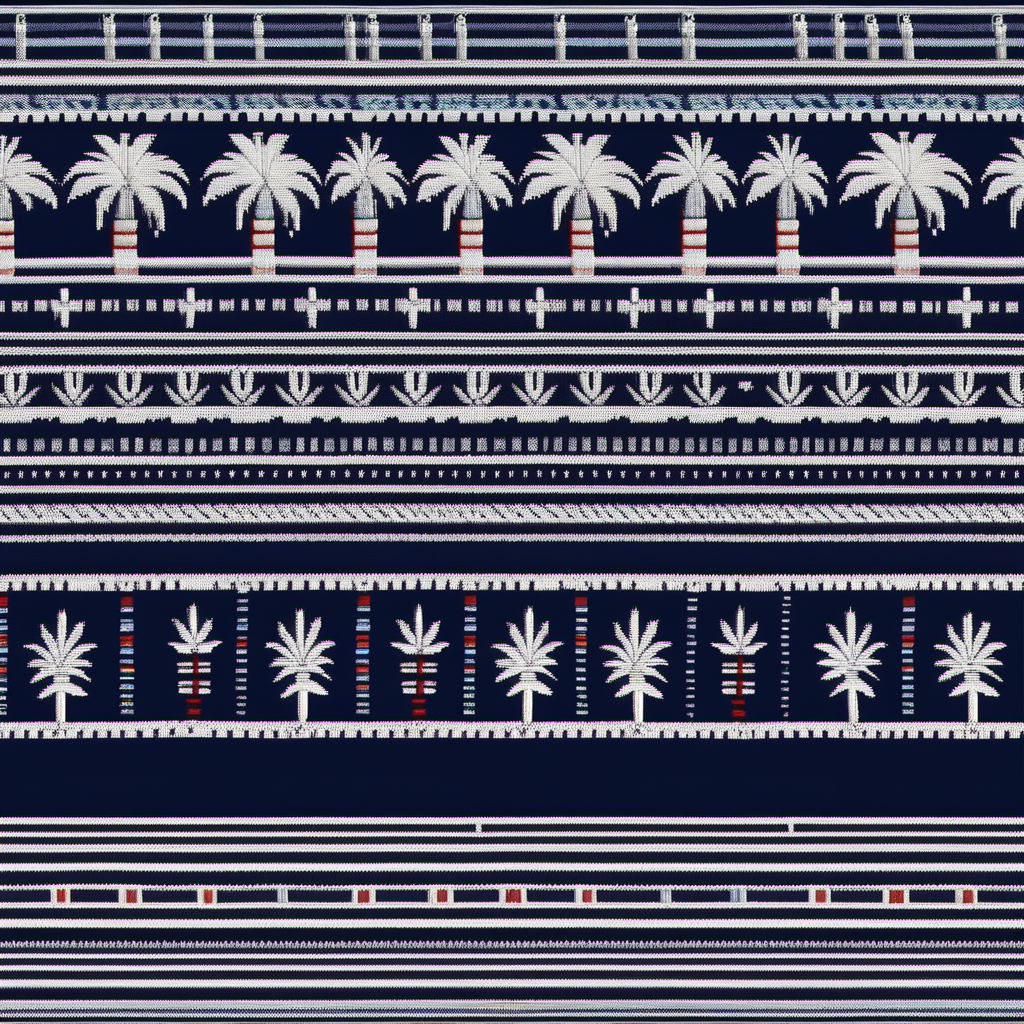}
    \caption{Al-Sadu generated image.}
    \label{fig:00036}
\end{figure}
\FloatBarrier

\begin{figure}[ht!]
    \centering
    \includegraphics[width=\columnwidth]{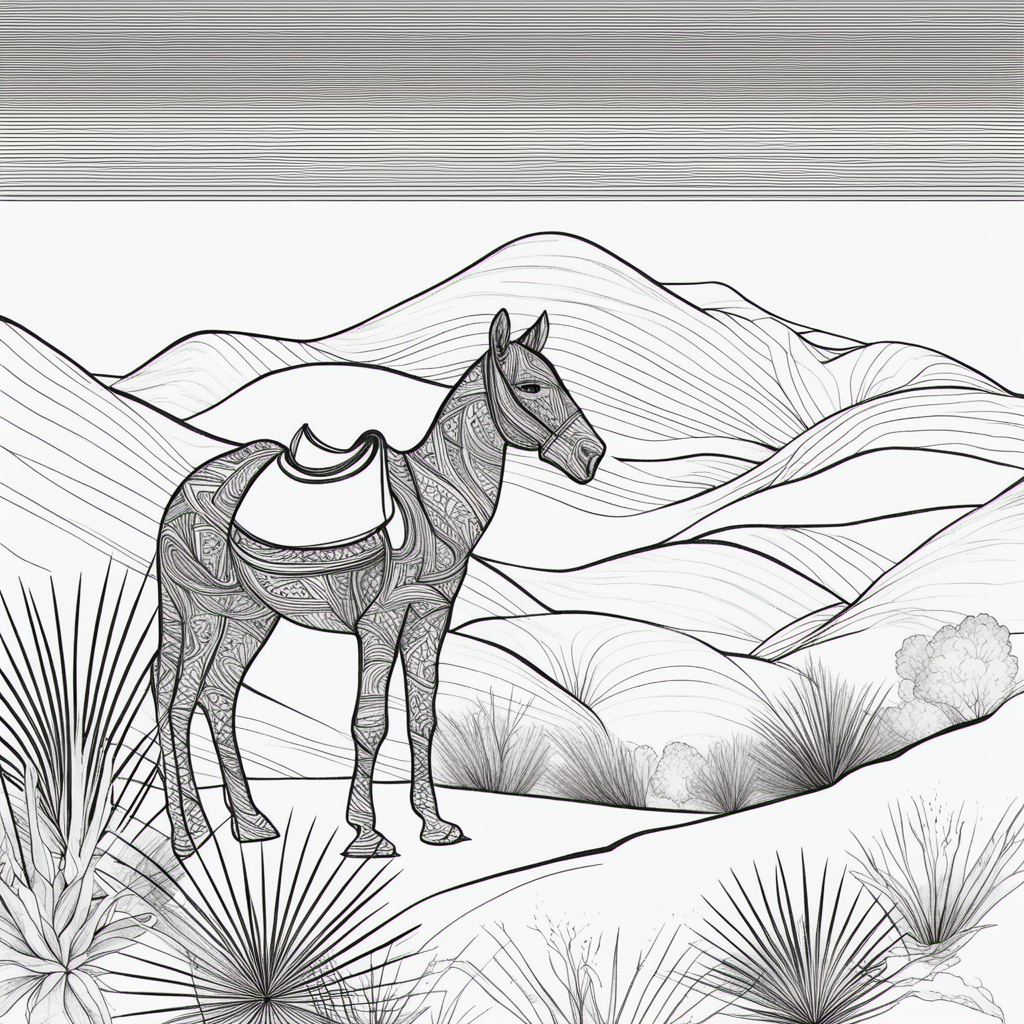}
    \caption{Detailed Horse in the desert generated image.}
    \label{fig:00061}
\end{figure}
\FloatBarrier

\begin{figure}[ht!]
    \centering
    \includegraphics[width=\columnwidth]{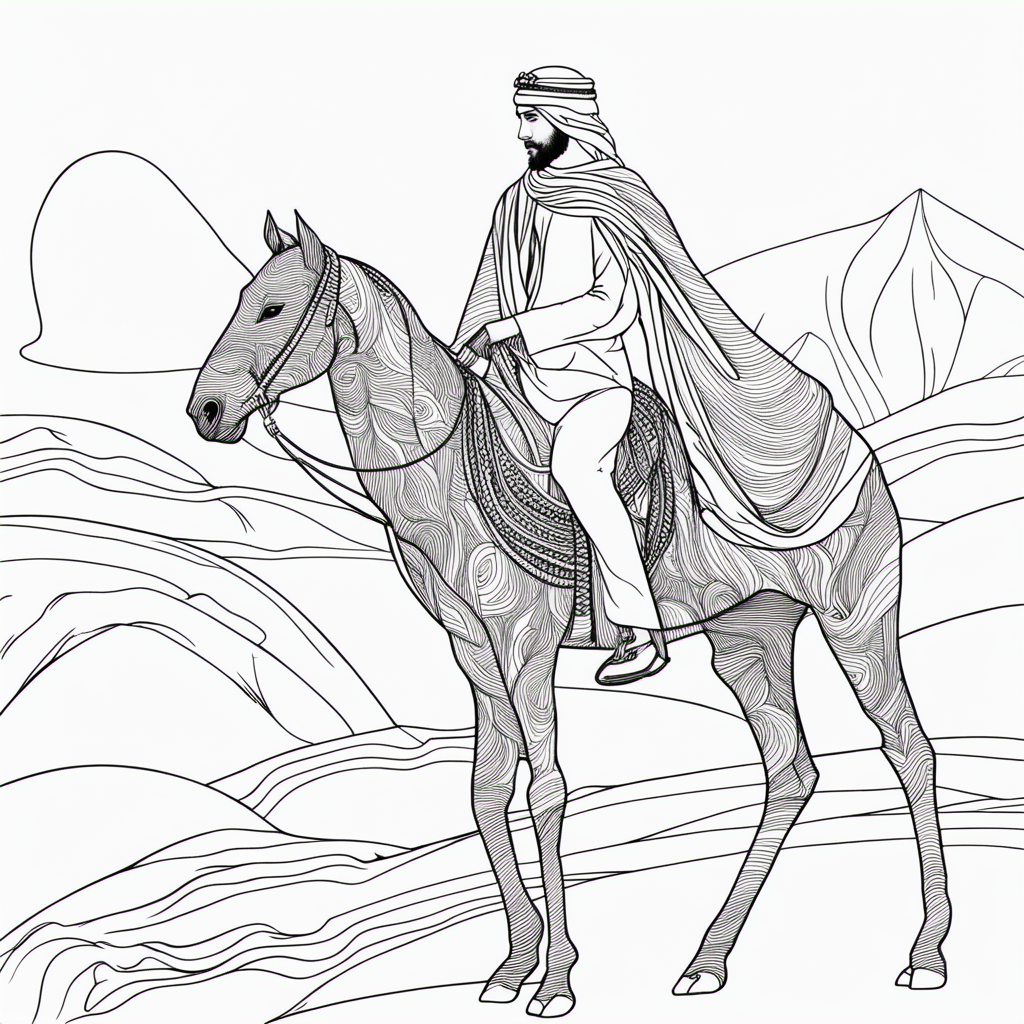}
    \caption{Emirati Man riding Horse in the desert generated image.}
    \label{fig:00063}
\end{figure}
\FloatBarrier

\begin{figure}[ht!]
    \centering
    \includegraphics[width=\columnwidth]{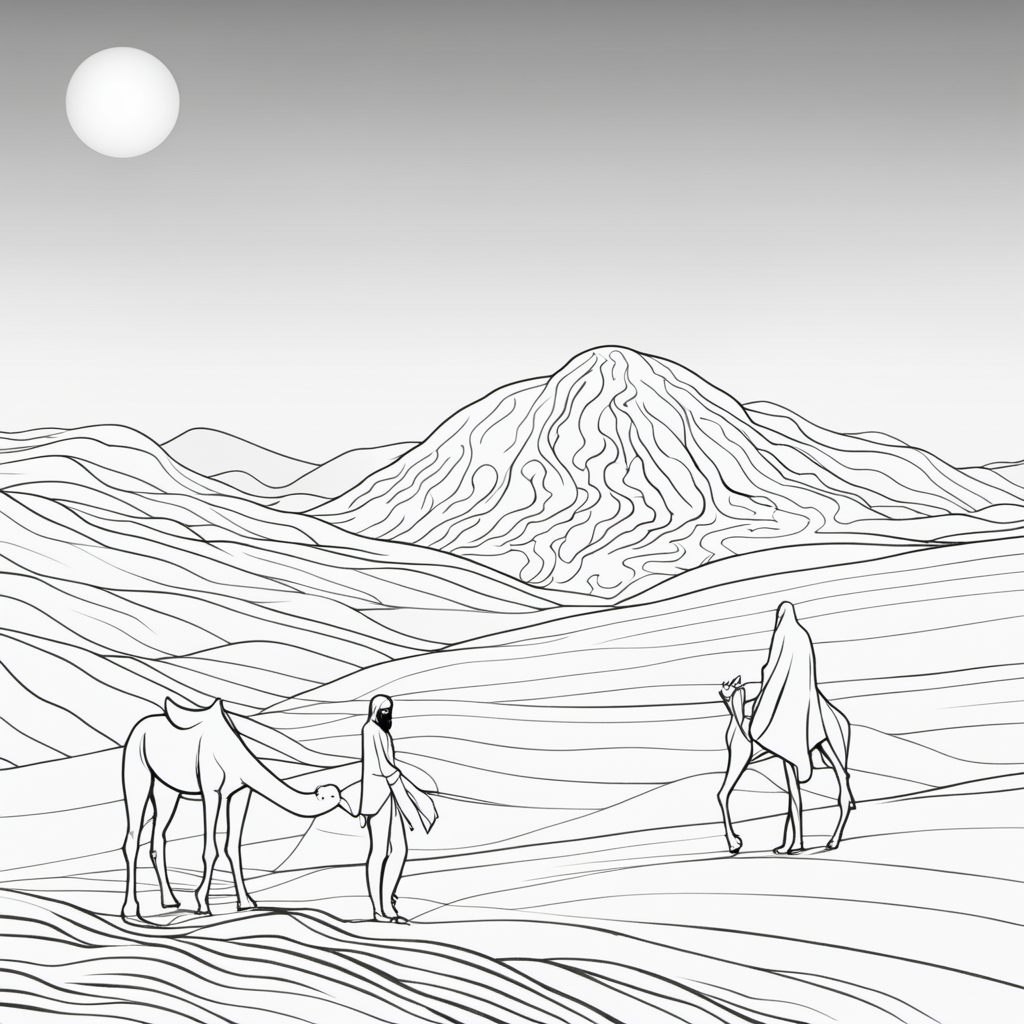}
    \caption{Sand dunes landscape generated image.}
    \label{fig:00064}
\end{figure}
\FloatBarrier

\begin{figure}[ht!]
    \centering
    \includegraphics[width=\columnwidth]{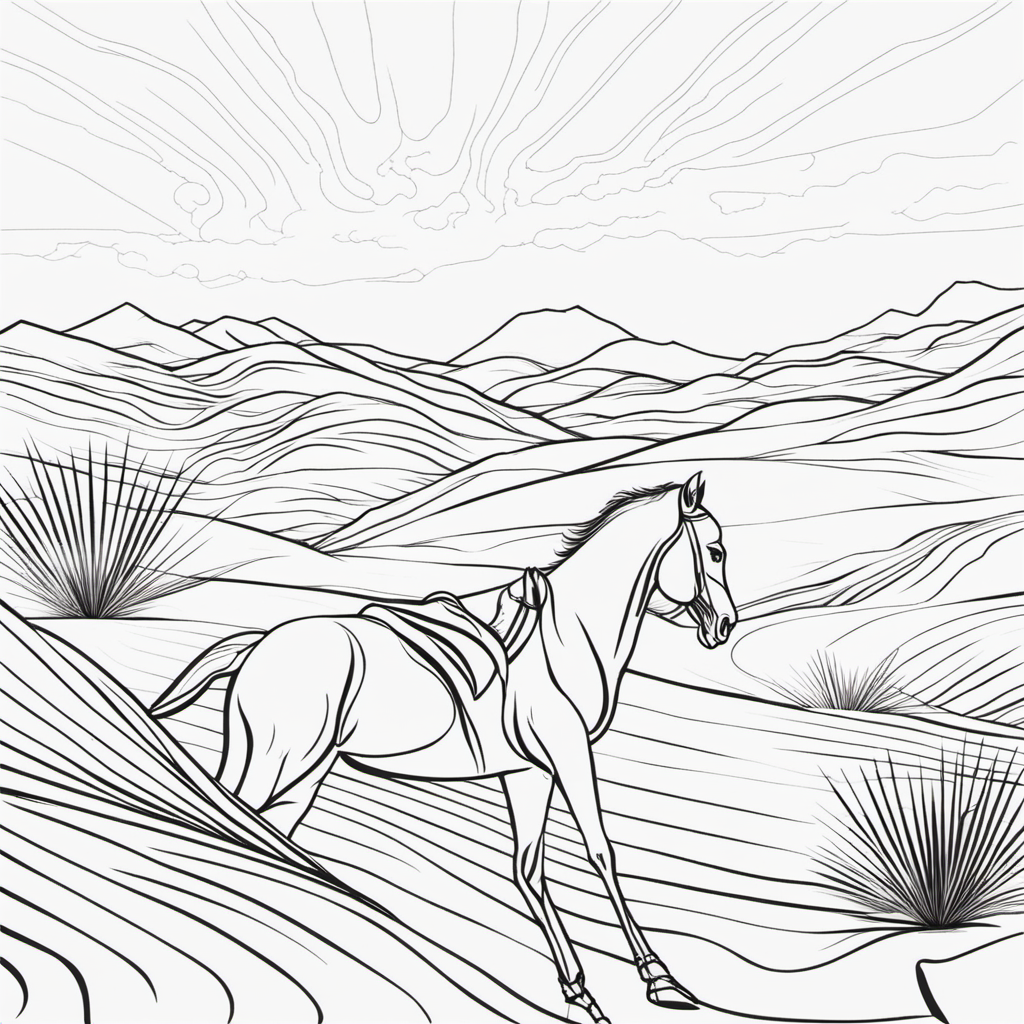}
    \caption{Sand dunes with Horse in the foreground generated image.}
    \label{fig:00068}
\end{figure}
\FloatBarrier

\begin{figure}[ht!]
    \centering
    \includegraphics[width=\columnwidth]{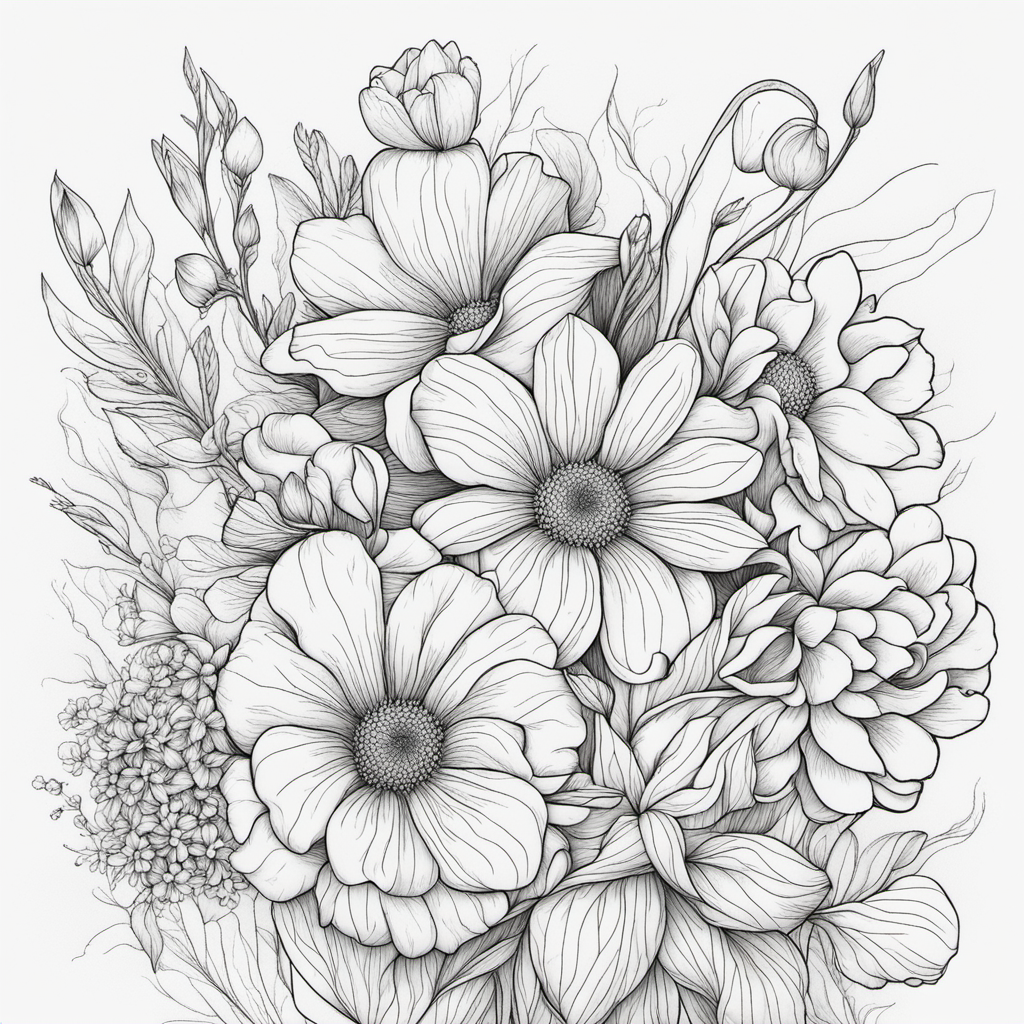}
    \caption{Detailed flower pattern generated image.}
    \label{fig:IMG_0044}
\end{figure}
\FloatBarrier

\bibliographystyle{IEEEtran}
\bibliography{IEEEabrv,Ref}

\begin{thebibliography}{10}
\providecommand{\url}[1]{#1}
\csname url@samestyle\endcsname
\providecommand{\newblock}{\relax}
\providecommand{\bibinfo}[2]{#2}
\providecommand{\BIBentrySTDinterwordspacing}{\spaceskip=0pt\relax}
\providecommand{\BIBentryALTinterwordstretchfactor}{4}
\providecommand{\BIBentryALTinterwordspacing}{\spaceskip=\fontdimen2\font plus
\BIBentryALTinterwordstretchfactor\fontdimen3\font minus \fontdimen4\font\relax}
\providecommand{\BIBforeignlanguage}[2]{{%
\expandafter\ifx\csname l@#1\endcsname\relax
\typeout{** WARNING: IEEEtran.bst: No hyphenation pattern has been}%
\typeout{** loaded for the language `#1'. Using the pattern for}%
\typeout{** the default language instead.}%
\else
\language=\csname l@#1\endcsname
\fi
#2}}
\providecommand{\BIBdecl}{\relax}
\BIBdecl

\bibitem{Samuel2022-iy}
B.~Samuel, H.~Wang, C.~Shi, Y.~Pan, Y.~Yu, W.~Zhu, and Z.~Jing, ``\BIBforeignlanguage{en}{The effects of coloring therapy on patients with generalized anxiety disorder},'' \emph{\BIBforeignlanguage{en}{Animal Model. Exp. Med.}}, vol.~5, no.~6, pp. 502--512, Dec. 2022.

\bibitem{charlson2019new}
F.~Charlson, M.~van Ommeren, A.~Flaxman, J.~Cornett, H.~Whiteford, and S.~Saxena, ``New who prevalence estimates of mental disorders in conflict settings: a systematic review and meta-analysis,'' \emph{The Lancet}, vol. 394, no. 10194, pp. 240--248, 2019.

\bibitem{krupnik2021depression}
V.~Krupnik, ``Depression as a failed anxiety: the continuum of precision-weighting dysregulation in affective disorders,'' \emph{Frontiers in Psychology}, vol.~12, p. 657738, 2021.

\bibitem{doi:10.1177/0276237420923290}
\BIBentryALTinterwordspacing
N.~Turturro and J.~E. Drake, ``Does coloring reduce anxiety? comparing the psychological and psychophysiological benefits of coloring versus drawing,'' \emph{Empirical Studies of the Arts}, vol.~40, no.~1, pp. 3--20, 2022. [Online]. Available: \url{https://doi.org/10.1177/0276237420923290}
\BIBentrySTDinterwordspacing

\bibitem{Al-Yateem2023-os}
N.~Al-Yateem, A.~M.~A. Lajam, M.~M.~G. Othman, M.~A.~A. Ahmed, S.~Ibrahim, A.~Halimi, F.~R. Ahmad, M.~A. Subu, J.~M. Dias, S.~A. Rahman, A.~R. Saifan, and H.~Hijazi, ``\BIBforeignlanguage{en}{The impact of cultural healthcare practices on children's health in the united arab emirates: a qualitative study of traditional remedies and implications},'' \emph{\BIBforeignlanguage{en}{Front. Public Health}}, vol.~11, p. 1266742, Oct. 2023.

\bibitem{IntangibleCulturalHeritage2019}
\BIBentryALTinterwordspacing
Dec 2019. [Online]. Available: \url{https://ich.unesco.org/en/USL/al-sadu-traditional-weaving-skills-in-the-united-arab-emirates-00517}
\BIBentrySTDinterwordspacing

\bibitem{Shouk_2023}
\BIBentryALTinterwordspacing
A.~A. Shouk, Dec 2023. [Online]. Available: \url{https://www.thenationalnews.com/uae/2023/12/02/uaes-52nd-union-day-celebration-combines-ancient-tradition-with-modern-technology/}
\BIBentrySTDinterwordspacing

\bibitem{Kriticka_2023}
A.~Kriticka, ``The effectiveness of chromotherapy on youth,'' \emph{Journal of Forensic Science and Research}, vol.~7, no.~1, p. 049–054, 2023.

\bibitem{buildings12020234}
\BIBentryALTinterwordspacing
C.~Jung, N.~S.~A. Mahmoud, G.~El~Samanoudy, and N.~Al~Qassimi, ``Evaluating the color preferences for elderly depression in the united arab emirates,'' \emph{Buildings}, vol.~12, no.~2, 2022. [Online]. Available: \url{https://www.mdpi.com/2075-5309/12/2/234}
\BIBentrySTDinterwordspacing

\bibitem{10.3389/fpsyg.2022.928048}
\BIBentryALTinterwordspacing
Y.~Li, X.~Li, Z.~Lou, and C.~Chen, ``Long short-term memory-based music analysis system for music therapy,'' \emph{Frontiers in Psychology}, vol.~13, 2022. [Online]. Available: \url{https://www.frontiersin.org/articles/10.3389/fpsyg.2022.928048}
\BIBentrySTDinterwordspacing

\bibitem{Santana_Lima_Torcate_Fonseca_Santos_2021}
\BIBentryALTinterwordspacing
M.~A.~d. Santana, C.~L.~d. Lima, A.~S. Torcate, F.~S. Fonseca, and W.~P.~d. Santos, ``Affective computing in the context of music therapy: a systematic review,'' \emph{Research, Society and Development}, vol.~10, no.~15, p. e392101522844, Nov. 2021. [Online]. Available: \url{https://rsdjournal.org/index.php/rsd/article/view/22844}
\BIBentrySTDinterwordspacing

\bibitem{10.3389/frai.2020.497864}
\BIBentryALTinterwordspacing
D.~Williams, V.~J. Hodge, and C.-Y. Wu, ``On the use of ai for generation of functional music to improve mental health,'' \emph{Frontiers in Artificial Intelligence}, vol.~3, 2020. [Online]. Available: \url{https://www.frontiersin.org/articles/10.3389/frai.2020.497864}
\BIBentrySTDinterwordspacing

\bibitem{9726576}
J.~Palsa, J.~Hurtuk, E.~Chovancova, and L.~Vaniscak, ``Emotion detection as a supportive tool in color therapy,'' in \emph{2021 19th International Conference on Emerging eLearning Technologies and Applications (ICETA)}, 2021, pp. 287--292.

\bibitem{101126scienceadh4451}
\BIBentryALTinterwordspacing
Z.~Epstein, A.~Hertzmann, the Investigators~of Human~Creativity, M.~Akten, H.~Farid, J.~Fjeld, M.~R. Frank, M.~Groh, L.~Herman, N.~Leach, R.~Mahari, A.~Pentland, O.~Russakovsky, H.~Schroeder, and A.~Smith, ``Art and the science of generative ai,'' \emph{Science}, vol. 380, no. 6650, pp. 1110--1111, 2023. [Online]. Available: \url{https://www.science.org/doi/abs/10.1126/science.adh4451}
\BIBentrySTDinterwordspacing

\bibitem{podell2023sdxl}
D.~Podell, Z.~English, K.~Lacey, A.~Blattmann, T.~Dockhorn, J.~Müller, J.~Penna, and R.~Rombach, ``Sdxl: Improving latent diffusion models for high-resolution image synthesis,'' 2023.

\bibitem{hu2021lora}
E.~J. Hu, Y.~Shen, P.~Wallis, Z.~Allen-Zhu, Y.~Li, S.~Wang, L.~Wang, and W.~Chen, ``Lora: Low-rank adaptation of large language models,'' 2021.

\bibitem{YubinIBM_Anaka_Ma_2023}
\BIBentryALTinterwordspacing
Y.~M. is~a designer, engineer. He has worked~for IBM, P.~Anaka, and Y.~Ma, ``How to train an sdxl lora (koyha with runpod),'' Sep 2023. [Online]. Available: \url{https://aituts.com/sdxl-lora/}
\BIBentrySTDinterwordspacing

\bibitem{li2022blip}
J.~Li, D.~Li, C.~Xiong, and S.~Hoi, ``Blip: Bootstrapping language-image pre-training for unified vision-language understanding and generation,'' 2022.

\bibitem{Ma_2023}
\BIBentryALTinterwordspacing
Y.~Ma, ``Stable diffusion lora training settings for koyha ss, explained,'' Sep 2023. [Online]. Available: \url{https://aituts.com/lora-training-settings/}
\BIBentrySTDinterwordspacing

\end{thebibliography}

\end{document}